\documentclass[aps,pra,superscriptaddress,showpacs,floatfix]{revtex4}
\usepackage{graphicx}
\usepackage{bm}
\usepackage{longtable}
\usepackage{epsfig}
\usepackage{amsmath}
\usepackage{amssymb}

\begin{document}

\title{Ion Acceleration by Short Chirped Laser Pulses}

\author{Jian-Xing Li}
\affiliation{Max Planck Institute for Nuclear Physics, Saupfercheckweg~1, 69029 Heidelberg, Germany}
\author{Benjamin Galow}
\affiliation{Ziegelh\"auser Landstra{\ss}e~19, 69120 Heidelberg, Germany}
\author{Christoph H. Keitel}
\affiliation{Max Planck Institute for Nuclear Physics, Saupfercheckweg~1, 69029 Heidelberg, Germany}
\author{Zolt\'an Harman}
\affiliation{Max Planck Institute for Nuclear Physics, Saupfercheckweg~1, 69029 Heidelberg, Germany}


\begin{abstract}
Direct laser acceleration of ions by short frequency-chirped laser pulses is investigated theoretically.
We demonstrate that intense beams of ions with a kinetic energy broadening of about 1$\%$ can be generated.
The chirping of the laser pulse allows the particles to gain kinetic energies of hundreds of MeVs, which is
required for hadron cancer therapy, from pulses of energies of the order of 100~J. It is shown that few-cycle chirped
pulses can accelerate ions more efficiently than long ones, i.e. higher ion kinetic energies are reached with the same
amount of total electromagnetic pulse energy.
\end{abstract}


\maketitle


\section{Introduction}

The interaction of intense laser pulses with solids has recently attracted considerable interest.
This is largely due to its potential application for accelerating charged particles
\cite{plasma2,plasma1,plasma3,Esirkepov2006prl,plasma4,plasma5,plasma6,plasma7,plasma8,plasma9,plasma10,plasma11,badziak,sal-prl2,crossed,salaminreview,DiP12,Peralta2013,Haberberger2012,Hooker2013,Zigler2013,zoltan2011pra}.
Tumor therapy with accelerated ion beams (see e.g.~\cite{cancer,debus1,med1,med2,med3,cancerrev}) would in particular benefit from
the replacement of conventional accelerators by all-optical devices, which may become compact and
inexpensive in future, allowing this form of cancer therapy to be accessible for more patients.

In this article, we demonstrate the feasibility of generating ion beams by shining an appropriately chirped short laser pulse on a target.
At sufficiently high laser intensities the electrons are quickly ionized, and the ions get directly accelerated by the laser field.
Modulating the frequency of the pulse leads to efficient particle energy gain from the field, as it was shown before~\cite{chirp-PRL,chirp0,chirp1,chirp2}.
Here we consider the case of short chirped pulses, when the time duration of the pulse is comparable to a single cycle. We found that acceleration
by such short pulses may be more efficient than by long chirped pulses, i.e. the same final ion kinetic energies can be reached with a lower
pulse energy.

\begin{figure}[th]
\begin{center}
\includegraphics[width = 0.8 \columnwidth]{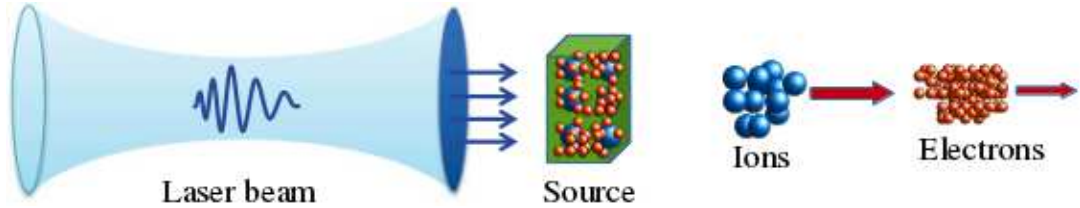}
\caption{\label{fig:scheme}
A schematic view of the studied laser acceleration set-up. A chirped few-cycle laser pulse is shot on a target. The ionized electrons
are accelerated fast in the forward direction, which is followed by the more inert ions, also directly accelerated by the laser pulse. Note that the transverse target size is magnified for illustration purpose. In the simulations it is much smaller than the focal radius of the laser beam.}
\end{center}
\end{figure}

The acceleration scheme is shown on Fig.~\ref{fig:scheme}. We assume a solid-density or underdense target consisting of carbon ions, typically, bare nuclei,
which can be generated via ionization by an intense pre-pulse. Collective plasma effects become important when the target thickness $L$ in propagation direction of the laser is longer than the
wave length of the induced plasma wave, i.e. $\lambda_p=2\pi c/\omega_p$ with plasma frequency $\omega_p^2=4\pi n_e e^2/m_e$. In the last formula $n_e$ is the electron density, $e$ the electron charge and $m_e$
the mass of an electron.
When interacting with the accelerating pulse, first the lighter electrons are accelerated, they are pushed in the
forward direction, as shown previously in Ref.~\cite{chirp-PRL}. The electrons are followed by the carbon ions, which are directly accelerated by the chirped
pulse.


\section{Laser acceleration simulations}

In order to access the efficiency of laser acceleration by means of few-cycle chirped pulses, we perform simulations based on the classical relativistic equations
of motion. In the following subsection, we provide a description of short laser pulses, based on earlier works~\cite{Yousef-fields,LiRR}. Subsequently, we describe
the particle dynamics in the presence of such intense chirped pulses.

\subsection{Description of few-cycle chirped focused laser pulses}

We consider a  circularly polarized laser field propagating along the $z$-direction. In analogy with Ref.~\cite{Esarey:95},
the vector potential for the focused ultra-short laser beam is represented as
$\bm{A}=\frac{E_0}{k_0}\left(\hat{x}  \psi (\mathbf{r}, \eta) +i\hat{y}\psi (\mathbf{r}, \eta)e^{i\pi/2} \right) e^{i\eta}$.
Here, the following notations have been introduced: the electric field amplitude is $E_0 = 4\sqrt{P/c}/w_0$, where $P$ is the peak power of a laser pulse,
$c$ stands for the light velocity in vacuum, and $w_0$ is the laser focal radius. The phase of the field is given as $\eta = \omega_0 t - k_0 z$.
The wave number is $k_0 = \omega_0/c$, where $\omega_0$ is the original (unchirped) frequency of laser at the focal point. When the chirping effect
is taken into account, the laser frequency and radius have to be modified as follows: $\omega = \omega_0 \left( 1+b\eta\right)$, and
$w = w_0/\left(1+b\eta\right)$, where $b$ is the dimensionless chirp parameter. In these formulas, we introduced the quantities
$\psi = f (1+i\eta/s^2) e^{i\phi_0-f \rho^2-\eta^2/(2s^2)}$, $f = i / ( i+\nu / z_r)$, $\nu = z + \eta /(2 k)$,  $\rho = r/w$,
$r = \sqrt{x^2+y^2}$, $s = \omega_0 \tau/2\sqrt{2\log{2}}$, and $\tau$ is pulse duration. Furthermore, $z_r = k w^2/2$ is the Rayleigh length, $k = \omega/c$, and
$\phi_0$ is a constant phase. When introducing the frequency modulation, the phase parameter $\eta$ and the
pulse length parameter $s$ do not need to be modified. $\hat{x}$ and $\hat{y}$ stand for unit vectors in the $x$ and $y$ directions orthogonal to
the lasers propagation direction.

Note that the temporal envelope of the laser beam is not factorized in this pulsed solution of the wave equation.
The scalar potential $\phi$ is thought to have a similar expression as that of the vector potential $\bm{A}$, and it can be calculated
from the Lorentz gauge condition $\partial \phi/\partial t + \nabla \cdot \bm{A}=0$.
The electromagnetic fields are derived from $\bm{E} = -\partial \bm{A}/\partial t - \nabla \phi, \bm{B}=\nabla\times\bm{A} $:
 $\bm{E}=\bm{E}^{(\hat{x})}+\bm{E}^{(\hat{y})}$,  $\bm{B}=\bm{B}^{(\hat{x})}+\bm{B}^{(\hat{y})}$, herewith,
\begin{eqnarray}
E_x^{(\hat{x})}&=&\frac{-E_1}{C_1^2} \left[C_1^3-f^2 C_1\frac{x^2}{z_r^2}+\frac{f C_1}{k z_r}-\frac{2 i f k x^2}{z_r \left(2 k C_2 +\eta\right)^2}\right] \,, \nonumber\\
E_y^{(\hat{x})}&=&\frac{E_1 f x y}{C_1^2 z_r^2}\left[f C_1+\frac{2 i k z_r}{(2 k j_z+\eta)^2}\right]\,,\nonumber\\
E_z^{(\hat{x})}&=&\frac{-E_1 f x}{C_1^2 z_r^2}\left[\frac{i f C_1}{k}-C_1 z_r \left(i+\frac{if^2 r^2}{4 z_r^2}-\frac{\eta}{s^2}\right)
  -\frac{i C_1 z_r}{s^2+i\eta}-z_r C_2\right] \,,\label{eq:3}\\
B_x^{(\hat{x})}&=&0 \,,\nonumber\\
B_y^{(\hat{x})}&=&E_1\left(\frac{i f}{2 k zr}-i-\frac{i f^2 r^2}{4 z_r^2}+\frac{\eta}{s^2}+\frac{1}{is^2-\eta}\right) \,,\nonumber\\
B_z^{(\hat{x})}&=&\frac{E_1 f y}{z_r}\,, \nonumber
\end{eqnarray}\\
and, furthermore,
\begin{eqnarray}
E_y^{(\hat{y})}&=&\frac{-E_1}{C_1^2} \left[C_1^3-f^2 C_1\frac{y^2}{z_r^2}+\frac{f C_1}{k z_r}-\frac{2 i f k y^2}{z_r \left(2 k C_2 +\eta\right)^2}\right] \,, \nonumber\\
E_x^{(\hat{y})}&=&\frac{E_1 f x y}{C_1^2 z_r^2}\left[f C_1+\frac{2 i k z_r}{(2 k j_z+\eta)^2}\right] \,,\nonumber\\
E_z^{(\hat{y})}&=&\frac{-E_1 f y}{C_1^2 z_r^2}\left[\frac{i f C_1}{k}-C_1 z_r \left(i+\frac{if^2 r^2}{4 z_r^2}-\frac{\eta}{s^2}\right)
  -\frac{i C_1 z_r}{s^2+i\eta}-z_r C_2\right] \,,\label{eq:3}\\
 B_y^{(\hat{y})}&=&0 \,,\nonumber\\
B_x^{(\hat{y})}&=&-E_1\left(\frac{i f}{2 k zr}-i-\frac{i f^2 r^2}{4 z_r^2}+\frac{\eta}{s^2}+\frac{1}{is^2-\eta}\right) \,,\nonumber\\
B_z^{(\hat{y})}&=&\frac{-E_1 f x}{z_r} \,, \nonumber
\end{eqnarray}
where $E_1=E_0\psi e^{i\left(\eta+b\eta^2\right)}/S_0$, with the normalization parameter $S_0 = [-s_1^2-1/(k z_r)]/s_1, s_1=i[1+1/(2 k z_r)+1/s^2]$, $j_z=z+iz_r$, and
the following short-hand notations have been introduced:
\begin{eqnarray}
C_1&=&i+\frac{i k^2 r^2-2 k j_z -\eta}{(2 k j_z +\eta)^2}+\frac{s^2+i\eta s^2-\eta^2}{s^2(\eta-is^2)}\,,\\
C_2&=&\frac{-1}{(2 k j_z+\eta)^2}+\frac{4 k j_z+2\eta-2 i k^2   r^2}{(2 k j_z+\eta)^3}
 +\frac{\eta^2-s^2-i \eta s^2}{(i \eta  s+s^3)^2}+\frac{2 i \eta +s^2}{i \eta s^2+s^4}\,.\nonumber
\end{eqnarray}
These field expressions are in concordance with the sub-cycle pulse field of Ref.~\cite{Lin_2006} and the long-pulse field of Ref.~\cite{Salamin_2002}.
Such fields were also employed recently to simulate quantum radiation reaction effects for an ensemble of electrons interacting with ultra-short pulses~\cite{LiRR}.
For the much slower ions studied here, the influence of radiation reaction can be neglected~\cite{crossed}.


\subsection{Particle dynamics}

The time-dependent dynamics of an ensemble of interacting ions is considered. An ion indexed by $j$, of mass $m$ and charge $q$ is accelerated to relativistic
energy and momentum, respectively, of ${\cal E}_j=\gamma_j mc^2$ and $\bm{p}_j=\gamma_j mc\bm{\beta}_j$, where $\bm{\beta}_j$ is the velocity
of the particle divided by $c$, the velocity of light in vacuum, and $\gamma=(1-\beta^2)^{-1/2}$ is the Lorentz factor, when interacting with the
time-dependent fields $\bm{E}$ and $\bm{B}$ of an laser pulse. Thus, the dynamics is described by the coupled Newton-Lorentz equations, given in SI (International
System of Units):
\begin{eqnarray}\label{motion_coul}
\frac{d\bm{p}_j}{dt} &=& q \left( \bm{E}(\bm{r}_j)+\bm{E}_{j}^{\text{\tiny{int.}}}+c\bm{\beta}_j\times\left(\bm{B}
(\bm{r}_j) \right) \right)\,, \\
\frac{d{\cal{E}}_j}{dt} &=& qc\bm{\beta}_j\cdot \left( \bm{E}(\bm{r}_j)+ \bm{E}_{j}^{\text{\tiny{int.}}} \right) \,.  \nonumber
\end{eqnarray}
The electric field of the inter-ionic interaction is approximated by
$\bm{E}_{j}^{\text{\tiny{int}}} = - \sum_{k \ne j} \nabla\phi_{jk}$, with the Coulomb interaction scalar potential
\begin{eqnarray}
\phi_{jk}&=&\frac{q}{4\pi\epsilon_0}\frac{1}{|\bm{r}_j-\bm{r}_k|}\label{Coulomb} \,.
\end{eqnarray}
Here, the relative displacement of two particles is $\bm{r}_{jk}=\bm{r}_j-\bm{r}_k$ and $\epsilon_0$ is the vacuum permittivity. The presence of
plasma electrons can be neglected when simulating the ions' acceleration dynamics, as they are blown off first by the pulse, as it was shown by
particle-in-cell simulations~\cite{chirp-PRL}.

Since the electromagnetic fields have a complex mathematical structure, and the inclusion of the ions' interaction leads to coupled motion, one has to solve
the above differential equations numerically. A numerical integration of Eqs.~(\ref{motion_coul}) yields the particles' velocities $\bm{\beta}_j$ and thus also
their energy gain at a final time equal to many laser cycles.

\section{Results and Discussion}

\begin{figure}[th]
\begin{center}
\includegraphics[width = 0.45 \columnwidth]{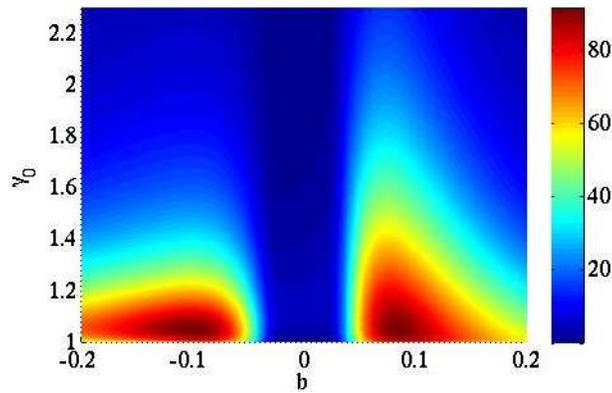}
\caption{
\label{fig:gain-vs-gamma0-vs-b}
Energy gain in units of MeV/u in dependence on the chirp parameter $b$ and the initial dimensionless energy parameter $\gamma_0$
(i.e. Lorentz factor, $E_0 = \gamma_0 m c^2$). The wavelength of the laser is chosen to be 1~$\mu$m and its power to 10~PW. The pulse duration is equal to
3 laser cycles, and the focal waist radius is equal to $\lambda$. A circularly polarized laser pulse is employed.}
\end{center}
\end{figure}

We present first results of calculations for a single test ion. Bare carbon nuclei (C${}^{6+}$ ions) are chosen here because of their higher contrast of dose deposition
in the tissue as compared to protons.
In order to find the optimal parameters of the pulse and the optimal initial velocity of the ion, we performed numerical simulations by calculating the energy gain
for a particle as a function of the chirp parameter $b$ and the initial energy $E_0 = \gamma_0 m c^2$ of the ion. The results
are presented on Fig.~\ref{fig:gain-vs-gamma0-vs-b}. The optimal chirping is in the range around $b=\pm 0.1$, with low initial kinetic energies around
$\gamma_0 \approx 1.05$. We note that for the circularly polarized pulses employed here, the dependence of the energy on the chirping parameter is a slowly-varying
function. This is not the case for linearly polarized fields, where this function shows a strong oscillatory behaviour~\cite{chirp-PRL}. Therefore, the appropriate
chirping can be more practically implemented in experiments. The maximal energy gain which can be reached in this setting is around 100 MeV/u, reaching the range of
interest in medical applications.

\begin{figure}[th]
\begin{center}
\includegraphics[width = 0.45 \columnwidth]{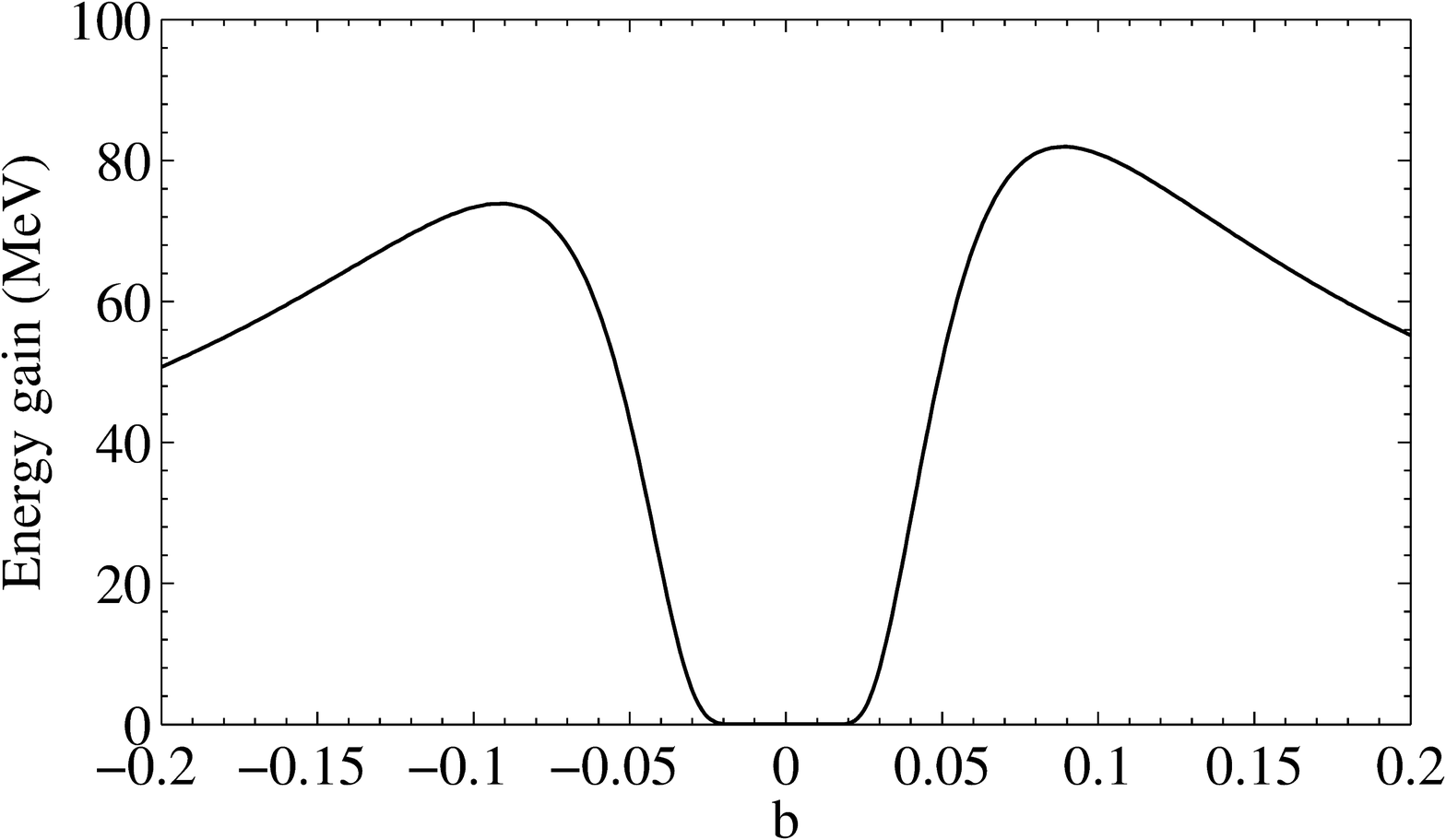}
\caption{
\label{fig:gain-vs-b}
Energy gain per nucleon for C${}^{6+}$ nuclei as a function of the chirp parameter $b$. The nuclei are assumed to be initially at rest, i.e. $\gamma_0=1$.
The other parameters are as described for Fig.~\ref{fig:gain-vs-gamma0-vs-b}.
}
\end{center}
\end{figure}

\begin{figure}[th]
\begin{center}
\includegraphics[width = 0.45 \columnwidth]{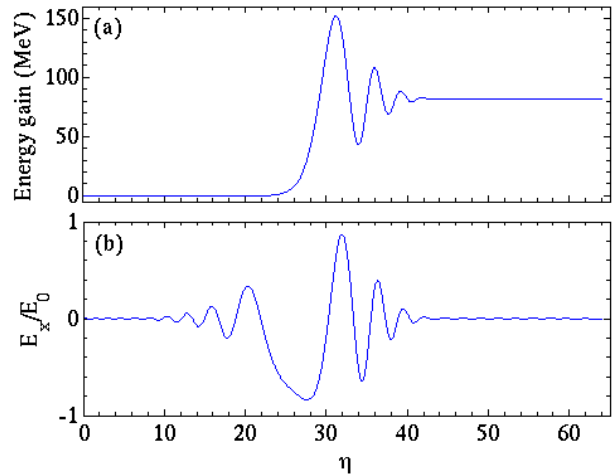}
\caption{
\label{fig:gainpulse}
(a) Energy gain per nucleon for C${}^{6+}$ nuclei and (b) the electric field of the pulse as a function of the phase $\eta$, for
a fixed optimal chirp parameter of $b = 0.089$. The other parameters are as described for Fig.~\ref{fig:gain-vs-gamma0-vs-b}.
}
\end{center}
\end{figure}

As the previous figure shows, the acceleration by a chirped short pulse is optimal when the initial kinetic energy of particles is low. Motivated by this,
we separately consider the experimentally advantageous situation when the particles are initially at rest, i.e. $\gamma_0 = 1$. The energy gain is shown on
Fig.~\ref{fig:gain-vs-b} in dependence of the chirp parameter $b$. The maximal gain that can be achieved by such pulses when the particles are at rest initially
is approx. 80~MeV/u. In the case of circularly polarized fields, similar gains can be reached when applying a negative frequency chirp ($b<0$), i.e. when the
carrier frequency of the laser decreases in time.
Additionally, circularly polarized laser beams have the advantage over linearly polarized beams that they are not sensitive to small variations of the chirp parameter $b$ - cf. Fig.~\ref{fig:gain-vs-b} with Fig.~2 of Ref. \cite{chirp-PRL}.
Fig.~\ref{fig:gainpulse} shows, for an optimal chirp parameter $b = 0.089$, the energy gain and the corresponding pulse as a function of the longitudinal
displacement $z$, confirming that the final kinetic energy around 100~MeV is reached on a sub-wavelength scale. The figure also illustrates --
in accordance with earlier findings~\cite{chirp-PRL} -- that it is the asymmetric part of the pulse, induced by the strong frequency modulation, which causes sudden acceleration of the ion.
This is in contrast to an unchirped laser pulse, where a charged particle would gain kinetic energy during the
first half of a laser cycle and subsequently loses this energy during the second half of the respective cycle due to the symmetry of the pulse.
However, we want to emphasize that frequency modulation over a large frequency bandwidth is presently experimentally accessible only at lower field intensities \cite{Goulielmakis:08, wirth}.

\begin{figure}[th]
\begin{center}
\includegraphics[width = 0.45 \columnwidth]{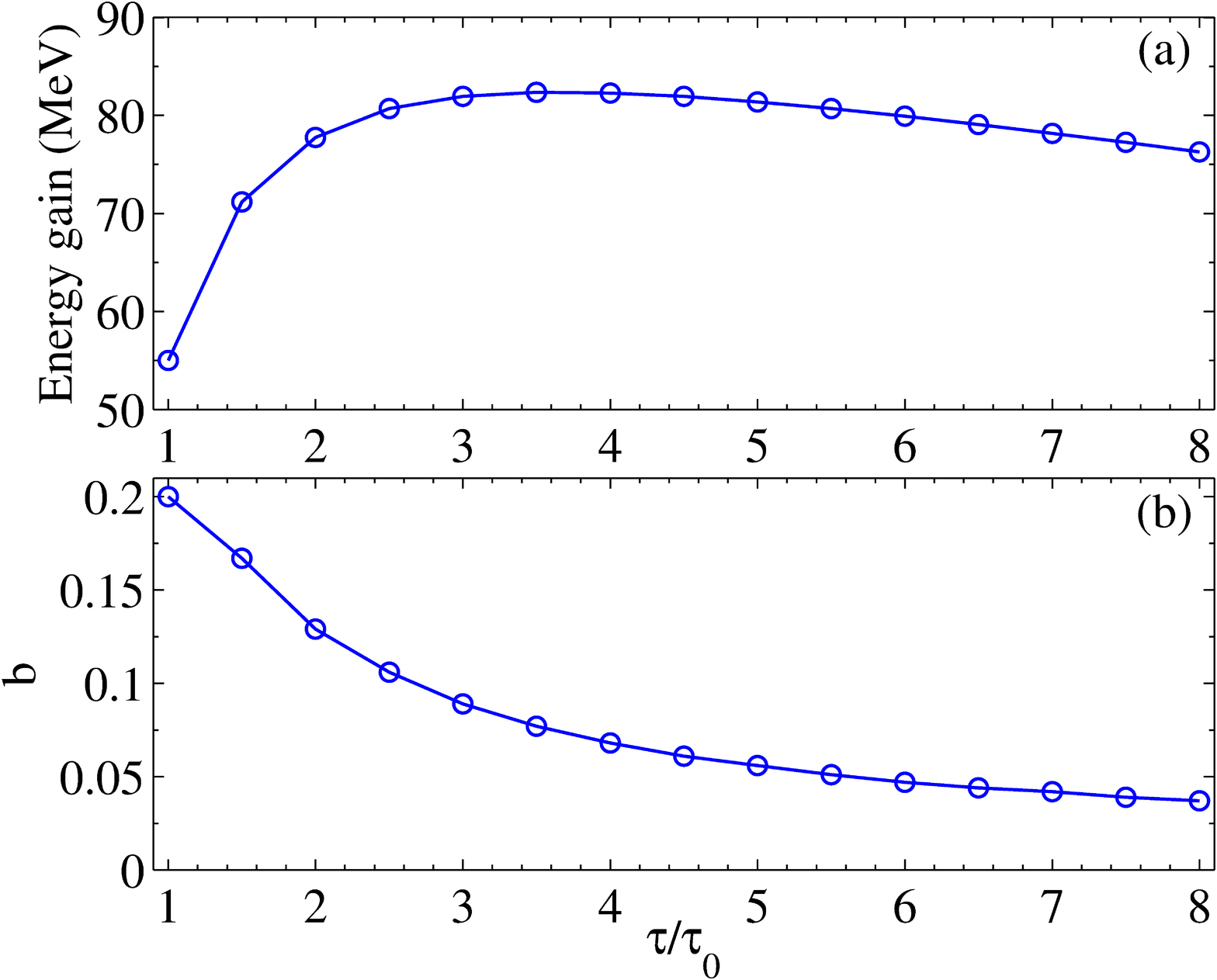}
\caption{
\label{fig:optimal}
(a) Energy gain per nucleon for C${}^{6+}$ nuclei and (b) and optimal chirp parameter $b$ as a function of the pulse duration $\tau$, given in units of the
period $\tau_0$. In these calculations, the pulse energy was kept fixed at a value of 100~J.
}
\end{center}
\end{figure}

In order to find the optimal pulse duration for acceleration, we performed single-particle simulations calculating the energy gain of a particle initially
at rest in dependence of this parameter, at a fixed total pulse energy. The results are displayed on Fig.~\ref{fig:optimal}~(a), showing that maximal
efficiency can be reached for pulses with a time duration of 3 to 4 cycles. Increasing the pulse duration can only decrease the final energy gain,
since the peak electric field of the pulse decreases if the energy of the pulse is distributed over a longer time. Furthermore, in pulses with durations below
3-4 periods, the ion does not spend sufficient time interacting with the field to reach its maximal velocity.
For each pulse duration, the chirp parameter $b$ was optimized independently. The corresponding values of the chirp parameters are displayed
on Fig~\ref{fig:optimal}~(b).

\begin{figure}[th]
\begin{center}
\includegraphics[width = 0.45 \columnwidth]{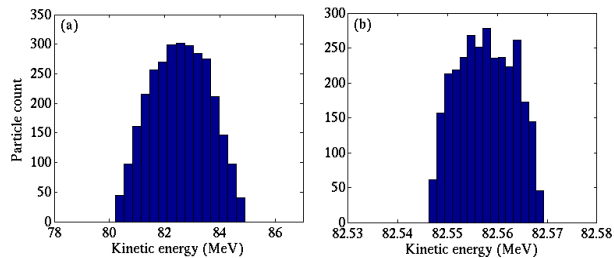}
\caption{
\label{fig:spread}
(a) Energy distribution histogram for an ensemble of 3000 interacting C$^{6+}$ ions, confined initially in a small cylinder (length 0.0004~$\mu$m, diameter: 0.002~$\mu$m).
(b) Energy distribution for the same set of ions, with the inter-particle Coulomb repulsion switched off. Note the different energy scales of the two sub-figures.
}
\end{center}
\end{figure}

\begin{figure}[th]
\begin{center}
\includegraphics[width = 0.45 \columnwidth]{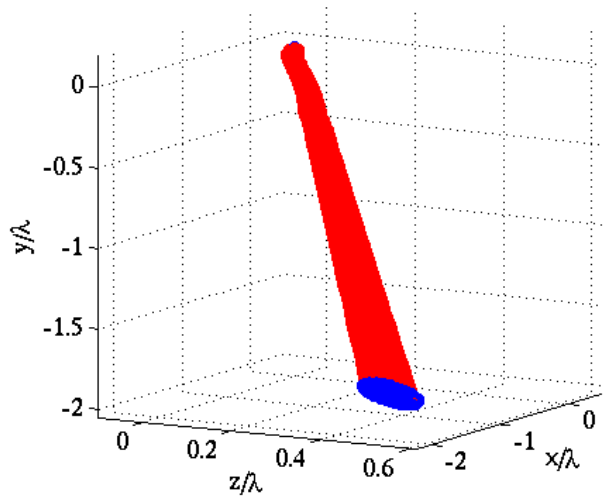}
\caption{
\label{fig:traj_overdense}
Trajectory of the overdense interacting C$^{6+}$ ion beam. Red curves are the trajectories of C$^{6+}$ ions, and blue points are the corresponding initial and final positions in the simulations with $-4s\leq\eta\leq 4s$. The parameters employed here are the same as those of Fig.~\ref{fig:spread}~(a).
}
\end{center}
\end{figure}

The energy distribution of a beam of ions generated by laser acceleration was determined by many-particle calculations based on the coupled Newton-Lorentz equations
[Eq.~(\ref{motion_coul})]. We assume that the ions are randomly distributed in a small nano-scale cylindrical volume before interacting with the short focused laser pulse.
The ions were assumed to be initially at rest, and have a solid-state density (2.4$\times 10^{24}$~ions/cm$^3$). Simulations were performed with and without the inclusion
of the ion-ion interaction to assess the effect of Coulomb repulsion on the energy gain and its broadening. In both cases, the set of random initial coordinates
utilized was kept constant. Results are shown on Fig.~\ref{fig:spread}. For this case collective plasma effects are negligible since the induced plasma wavelength, $\lambda_p = 0.0088\mu$m, is much larger than the target thickness, $L = 0.0004\mu$m. Including the particle-particle repulsion yields an order-of-magnitude
broader distribution than the simulation with artificially switching off the interaction, however, the average kinetic energy is not influenced by this:
in the first, realistic simulation, the average gain is 82.557~MeV/u, with a standard deviation of 1.042 MeV/u (1.262\%); in the latter case, the average gain
is the same, while the standard deviation is 0.006 MeV/u (0.007\%). Thus, we may conclude that for dense targets, it is indeed necessary to include inter-particle
interactions in realistic simulations, even at the very high laser intensities considered. The beam energy spread is in the medically applicable range, i.e.
approximately 1~\%, enabled by the small size of the target. Trajectory of the overdense interacting C$^{6+}$ ion beam is illustrated in Fig.~\ref{fig:traj_overdense}, and the emission angle is approximately 0.05 (radian), i.e. 2.865 (degree). Please note that the angle employs the unit of radian in the rest content. For the same conditions except taking a larger chirp parameter, $b = 0.1$, the emission angle increases from 0.05 to 0.051. And, for a longer laser pulse with the pulse duration $\tau = 4\tau_0$, the emission angle decreases to 0.048.

\begin{figure}[th]
\begin{center}
\includegraphics[width = 0.45 \columnwidth]{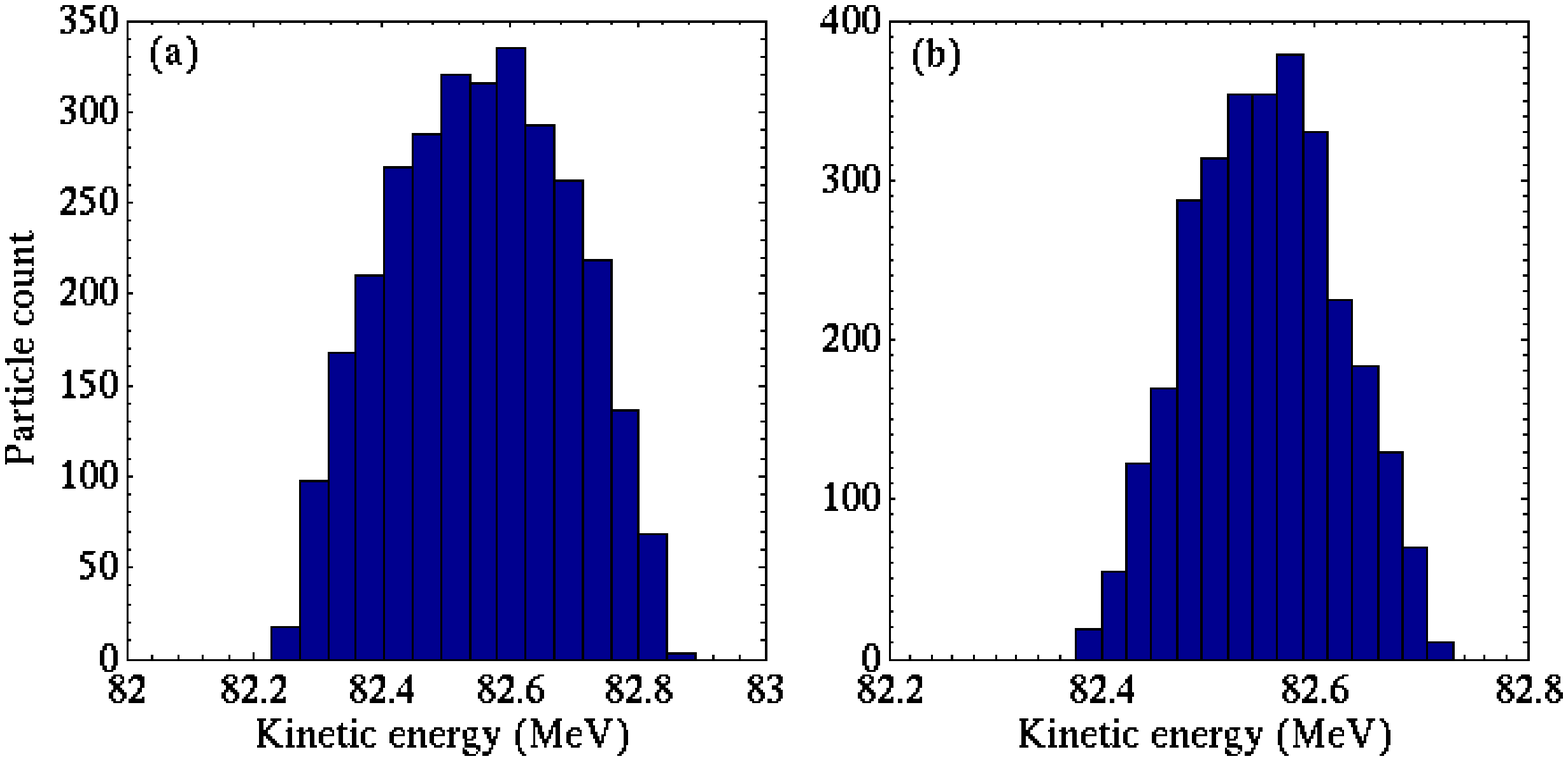}
\caption{
\label{fig:underdense}
(a) Energy distribution histogram for the case of an underdense target formed by 3000 interacting C$^{6+}$ ions,
confined initially in a cylinder (length 0.04~$\mu$m, diameter: 0.02$\mu$m)
(b) Energy distribution for the same underdense target, with the inter-particle interaction neglected.
}
\end{center}
\end{figure}

\begin{figure}[th]
\begin{center}
\includegraphics[width = 0.45 \columnwidth]{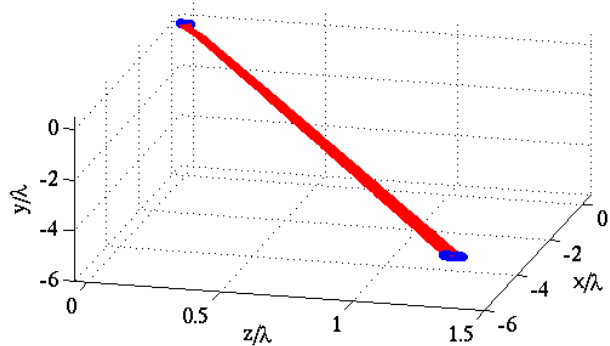}
\caption{
\label{fig:traj_underdense}
Trajectory of the underdense interacting C$^{6+}$ ion beam. Red curves are the trajectories of C$^{6+}$ ions, and blue points are the corresponding initial and final positions in the simulations with $-4s\leq\eta\leq 12s$. The parameters employed here are the same as those of Fig.~\ref{fig:underdense}~(a).
}
\end{center}
\end{figure}

Besides a nano-scale target, simulations were made for a more extended underdense carbon plasma target, which may be realized as an expanding plasma created
by a pre-pulse. The kinetic energy distribution for a target of ions with an initial density of 2.4$\times 10^{20}$~ions/cm$^3$ is presented as a histogram on
Fig.~\ref{fig:underdense}. The induced plasma wavelength, $\lambda_p = 0.88\mu$m, and the target thickness, $L = 0.04\mu$m. Therefore, collective plasma effects are ignored. In this case, particle-particle interactions are weaker, and thus the results are less sensitive to the presence of interaction.
We obtain an average ion gain of 82.553~MeV/u, with a standard deviation of 0.138~MeV/u (0.167\%) for the more realistic simulation with the ions' mutual repulsion
taken into account, and to the same average gain with a spread of 0.069~MeV/u (0.083\%) with the interaction neglected. The energies reached here
coincide with the results for a solid-density ion plasma target, however, the energy broadening is somewhat better. Trajectory of the underdense interacting C$^{6+}$ ion beam is illustrated in Fig.~\ref{fig:traj_underdense}. Emission angle is approximately 0.005 and 10 time smaller than that of the overdense case.

\section{Conclusions}

Relativistic many-particle simulations were performed in order to access the applicability of few-cycle chirped laser pulses to ion acceleration
for the purpose of hadron cancer therapy. An accurate description of the temporal and spatial structure of the laser fields was employed.
We have found that such pulses with durations of 3-4 cycles, when focused on small underdense or solid-density plasma targets, can produce ion beams
with properties in the range of medical requirements. They are also more efficient in accelerating ions than longer chirped pulses investigated in our
earlier work~\cite{chirp-PRL}: a further increase of pulse duration beyond the optimal value decreases the particles' energy gain, as the pulse energy
distributed over longer times leads to weaker accelerating fields.


\section*{Acknowledgments}

The authors acknowledge insightful and refreshing conversations with Yousef I. Salamin. J.-X. Li partially supported by National Natural Science Foundation of China (Grant No. 11304077).


\section*{Conflicts of Interest}

The authors declare no conflict of interest.

\bibliography{lit}

\end{document}